\documentclass[aps,prb,reprint,amssymb,amsmath,superscriptaddress]{revtex4-1}

\usepackage{mathtext}
\usepackage{amssymb}
\usepackage{graphicx}
\usepackage{amsmath}
\begin{document}

\title{Interaction correction to conductivity of Al$_x$Ga$_{1-x}$As/GaAs double quantum
well heterostructures near the balance}

\author{G.~M.~Minkov}
\affiliation{Institute of Metal Physics RAS, 620990 Ekaterinburg,
Russia}

\affiliation{Ural State University, 620000 Ekaterinburg, Russia}

\author{A.~V.~Germanenko}

\author{O.~E.~Rut}
\affiliation{Ural State University, 620000 Ekaterinburg, Russia}

\author{A.~A.~Sherstobitov}
\affiliation{Institute of Metal Physics RAS, 620990 Ekaterinburg,
Russia}

\affiliation{Ural State University, 620000 Ekaterinburg, Russia}

\author{A.~K.~Bakarov}
\author{D.~V.~Dmitriev}

\affiliation{Institute of Semiconductor Physics RAS, 630090
Novosibirsk, Russia}

\date{\today}

\begin{abstract}

The electron-electron interaction quantum correction  to the
conductivity of the gated double well Al$_x$Ga$_{1-x}$As/GaAs
structures is investigated experimentally. The analysis of the
temperature and magnetic field dependences of the conductivity tensor
allows us to obtain reliably the diffusion part of the interaction
correction for the regimes when the structure is balanced and when only
one quantum well is occupied. The surprising result is that the
interaction correction does not reveal resonant behavior; it is
practically the same for both regimes.

\end{abstract} 

\maketitle

\section{Introduction}
\label{sec:intr}

The double quantum well (DQW) structures exhibit a number of salient
features. For instance, the resistance of the structures with different
mobilities in the wells strongly depends on the potential profile of
the quantum wells and has a peak when the latter is
symmetric.\cite{Palevski90,Berk94,Slutzky96} The DQW systems are
convenient system to study a wide variety of the oscillatory phenomena
originated from the peculiarities of the Landau quantization of the
energy spectrum.\cite{Harrf97,Fletcher05,Duarte07,Mamani08,Mamani09}
The quantum corrections to the conductivity are also expected to
demonstrate peculiar behavior when the population of the quantum wells
and/or the interwell transition rate is varied. The interference
quantum correction in DWQ's is studied in
Refs.~\onlinecite{Averkiev98,Raichev00,Min00-3,Pagnossin08,Minkov10-1},
where the specific features of the interference induced
magnetoresistance and dephasing processes are investigated both
theoretically and experimentally. The correction due to the
electron-electron ({\it e-e}) interaction is investigated significantly
less.\cite{Pagnossin08,Arapov09}

The contribution of {\it e-e} interaction to the conductivity is
determined by two terms: singlet and multiplet. Singlet term does not
depend on interaction constant and it favors localization, i.e., it
leads to the conductivity  decrease with lowering temperature.  In
contrast, the contribution of the interaction in the  multiplet channel
depends  on the interaction constant $F_0^\sigma$ [or $\gamma_2=-
F_0^\sigma/(1+ F_0^\sigma)$] and gives antilocalization correction for
ordinary semiconductor structures. For diffusion regime ($T\tau\ll 1$,
where $\tau$ is the transport relaxation time), the interaction
correction to the conductivity is as follows\cite{Finkelstein83}
\begin{eqnarray}
 \delta\sigma_{ee}&=&K_{ee}G_0\ln(T\tau),\nonumber \\
 K_{ee}&=&1+(4n_v^2-1)\left[1-\frac{1+\gamma_2}{\gamma_2}\ln\left(1+\gamma_2\right)\right],
 \label{eq10}
\end{eqnarray}
where $G_0=e^2/\pi h\simeq 1.23\times10^{-5}$~Ohm$^{-1}$, $n_v$ is the
number of valleys. This correction is independent of the magnetic
field, while the Zeeman splitting is sufficiently small, $\textsl{g}
\mu_B B\ll T$, where $\textsl{g}$ is the Land\'{e} $\textsl{g}$-factor.
The first and second terms correspond to the contributions in the
singlet and multiplet channel, respectively.

Let us demonstrate how strongly the coefficient $K_{ee}$ in front of
the logarithm in Eq.~(\ref{eq10})  (just its value is determined
experimentally) depends on the valley degeneracy. For the 2D gas with
the single valley spectrum, $n_v=1$, at $\gamma_2=0.28$, that
corresponds to the electron density $n\simeq 10^{12}$ cm$^{-2}$ for
GaAs according to Ref.~\onlinecite{Zala01}, the multiplet contribution
is less than the singlet one and $K_{ee}=0.61$.  For the case of the
two valleys electron spectrum, as it takes place in [100] Si-MOS 2D
structures, the correction in the multiplet channel is larger than that
in singlet one and the coefficient $K_{ee}$ for the same electron
density should have opposite sign, $K_{ee}= -0.93$, i.e., the
correction should be antilocalizing.\footnote{This estimate corresponds
to the case when the valleys splitting and inter-valleys transitions
are absent and, therefore, only one interaction constant $F_0^\sigma$
determines the correction value}

It seems that double quantum well heterostructures based on the single
valley semiconductors should demonstrate analogous behavior. The
crucial change of the interaction contribution to the conductivity
should be observed when changing the relation between the electron
densities in the wells with the help of gate electrode, for example. If
one naively supposes that the total correction is  the sum of the
interaction contributions from each of the wells,  the value of
$K_{ee}$ about $1.07$ should be observed for particular case of the
different but close electron densities in the wells, $n_1\approx n_2
\approx 5\times 10^{11}$~cm$^{-2}$. When the electron density is the
same in the wells, $n_1=n_2 = 5\times 10^{11}$~cm$^{-2}$, we should
deal with the analog of the two valleys structure, for which the
coefficient in the multiplet term becomes equal to $15$, the
interaction contribution becomes antilocalizing with $K_{ee}=-0.91$.
But unlike the Si-MOS structure, for this case one can easily change
the relation between the singlet and multiplet contributions by varying
the densities ratio in the wells. Of course, such the giant change of
$K_{ee}$ can be observed in special structures only. Namely, the
electron densities and mobilities in the wells should be close.
Moreover, the scatterers should be common for the carriers in the
different wells in the sense that each specific impurity should scatter
the carriers of the lower and upper wells identically. In addition to
that, the interwell distance $d$ should be small, $\varkappa d<1$,
where $\varkappa$ is the inverse screening length, but the interwell
transition time $t_{12}$ should be large, $t_{12}\gg 1/T$. In reality,
it is very difficult to fulfil (and especially to check the fulfillment
of) all these requirements. However, because the qualitative
speculation presented above predicts very huge effect, it seems that
significant change of the interaction correction in double well
structure at varying of the density  should be observed easily even in
structures that fall short of this ideal. To the best of our knowledge,
such renormalization  of interaction contribution to the conductivity
in the singlet and multiplet channels at varying ratio of densities in
the wells was never observed experimentally. In this paper we try to
detect the resonant change of interaction contribution to the
conductivity in the double well structures.

\section{Experimental details}
\label{sec:exp}

The results presented in this paper have been obtained for just the
same samples, for which the weak localization effect has been
investigated in Ref.~\onlinecite{Minkov10-1}. The gated samples were
made on the basis of  the double quantum well heterostructures, in
which the two GaAs quantum wells of width $8$~nm are separated by
$10$~nm Al$_{0.3}$Ga$_{0.7}$As barrier. Two $\delta$ layers of Si have
been situated in the barriers to deliver the electrons in the wells.
The main doping $\delta$ layer of Si is in the center of barrier
separating the wells. To compensate the electric field of the Schottky
barrier, the second $\delta$ layer is located above the upper quantum
well at distance of $18$~nm from the well interface. Two
heterostructures, 3243 and 3154, distinguishing by the doping level
have been investigated. The main parameters of the samples for two
regimes considered in this paper are listed in Table~\ref{tab1}. The
regime when only the lower quantum well is occupied is referred as SQW
regime. Balance is the regime of the equal electron densities in the
wells.

The results obtained were mostly analogous and we will discuss in more
detail the results obtained for the structure 3243.

\begin{table}
\caption{Parameters of the structures investigated} \label{tab1}
\begin{ruledtabular}
\begin{tabular}{ccccc}
Structure &\multicolumn{2}{c}{\#3243}&\multicolumn{2}{c}{\#3154}\\
 Regime & SQW & balance & SQW & balance\\
  $V_g$ (V) & $-4.1$ & $-1.5$ & -3.6 & -2.0\\
 \colrule
 $n$ ($10^{11}$~cm$^{-2}$)\footnote{The electron density per quantum well.}
  & 7.0 & $7.5$ & 4.0 &$4.5$\\
 $\mu$ (10$^3$~cm$^{2}$/Vs) & 14.5&  $15$& 4.8 &$6.5$\\
 $K_{ee}\pm 0.05$, exp. &$0.60$& $0.57$ &$0.53$&$0.50$\\
  $K_{ee}$, theor. &$0.59$& $0.72$ &$0.52$& $0.59$\\
 \end{tabular}
\end{ruledtabular}
\end{table}

\section{Results and discussion}
\label{sec:results}

To extract the interaction contribution to the conductivity we have
used the unique property of this correction in the diffusive regime:
the interaction gives contribution to the one component of the
conductivity tensor, namely, to $\sigma_{xx}$, whereas
$\delta\sigma_{xy}=0$. At low interwell transition rate the components
of the conductivity tensor in the double well structures are simply the
sum of the components of each well,
$\sigma_{xx,xy}=\sigma^{(1)}_{xx,xy}+\sigma^{(2)}_{xx,xy}$. Therefore
the temperature dependence of $\sigma_{xx}$ at the high magnetic field,
$B>B_{tr}$ (where $B_{tr}=\hbar/2el^2$ is the transport magnetic field,
$l$ is  the transport mean free path), when the temperature dependence
of the WL correction is mainly suppressed, is determined by the
interaction correction only. This dependence should be logarithmic and
the slope of $\sigma_{xx}$~vs~$\ln{T}$ dependence should give the value
of $K_{ee}$.

The situation becomes more complicated at $T\tau> 0.1$, when the
ballistic contribution of interaction becomes important.\cite{Zala01}
This contribution results in the temperature dependent correction to
the mobility.\cite{Zala01,Min03-1} In its turn this leads to appearance
of the magnetic field dependence of $\Delta\sigma_{xx}=\sigma_{xx}(T)-
\sigma_{xx}(T_0)$ and the temperature dependence of
$\Delta\sigma_{xy}$:
\begin{eqnarray}
  \Delta\sigma_{xx}(B,T)&=&\sum_{i=1}^2\frac{1-\mu_i^2(T_0)B^2}{\left[1+\mu_i^2(T_0)B^2\right]^2}
   en_i\Delta\mu_i(T)   \nonumber\\
   &+ &\left[K_{ee}^{(1)}+K_{ee}^{(2)}\right]G_0\ln{\left(\frac{T}{T_0}\right)},  \label{eq20}\\
   \Delta\sigma_{xy}(B,T)&=&\sum_{i=1}^2\frac{2\mu_i(T_0)B}{\left[1+\mu_i^2(T_0)B^2\right]^2}
   en_i\Delta\mu_i(T),
  \label{eq25}
\end{eqnarray}
where $\Delta\mu_i(T)=\mu_i(T)-\mu_i(T_0)$. As seen from
Eq.~(\ref{eq20}) the temperature dependence of $\sigma_{xx}$ in this
case is determined not only by $K_{ee}^{(1)}$ and  $K_{ee}^{(2)}$, but
by $\mu_1$, $\mu_2$, $\Delta\mu_1$, and $\Delta\mu_2$ also. Things will
get better preferably when the mobilities in the wells are close to
each other. Then, as seen from Eq.~(\ref{eq20}), the temperature
dependence of $\sigma_{xx}$ at $B=1/\mu$ is determined by diffusion
interaction correction only. Therefore, let us start analysis of the
data from this case.

\begin{figure}
\includegraphics[width=\linewidth,clip=true]{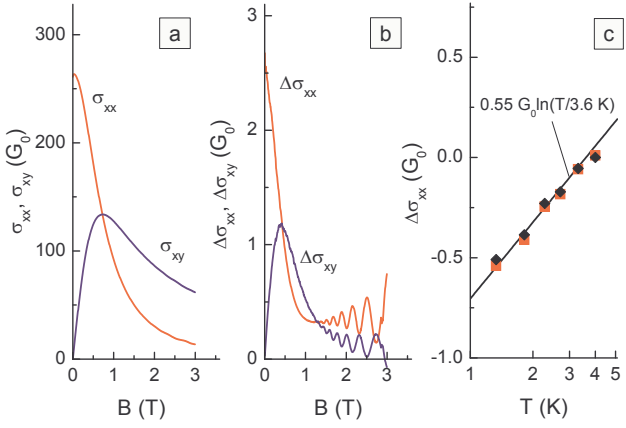}
\caption{(Color online) (a) The magnetic field dependences of $\sigma_{xx}$ and
$\sigma_{xy}$ taken at  $T=1.35$~K. (b)
The magnetic field dependences of $\Delta\sigma_{xx,xy}=\sigma_{xx,xy}(4.2\text{ K})-
\sigma_{xx,xy}(1.35\text{ K})$. (c) The temperature dependences of
$\Delta\sigma_{xx}=\sigma_{xx}(4.2\text{ K})-
\sigma_{xx}(T)$ at $B=1/\mu = 0.67$~T (squares) and that found from the Hall effect as
described in text  (diamonds).  $V_g=-1.5$~V.
}\label{f1}
\end{figure}

Detailed analysis of the gate voltage dependencies of the electron
densities and mobilities performed for these structures in
Ref.~\onlinecite{Minkov10-1} shows that the close values of the
mobility in the wells occur near the balance, $n_1=n_2$, which happens
for the structures 3243 and 3154 at $V_g=-1.5$~V and $V_g=-2$~V,
respectively. The magnetic field dependences of $\sigma_{xx}$,
$\sigma_{xy}$ and $\Delta\sigma_{xx,xy}=\sigma_{xx,xy}(4.2\text{ K})-
\sigma_{xx,xy}(1.35\text{ K})$ taken for the structure 3243 at
$V_g=-1.5$~V are presented in Fig.~\ref{f1}(a) and Fig.~\ref{f1}(b),
respectively. One can see that $\Delta\sigma_{xx}$ decreases strongly
up to $B=1$~T therewith $\Delta\sigma_{xy}$ is not small over the whole
magnetic field range. (The value of $B_{tr}$ at this gate voltage is
about $10^{-2}$~T so the variation of $\sigma_{xx}$ and $\sigma_{xy}$
does not associated with the contribution of the weak localization
correction). Such variations of $\Delta\sigma_{xx}$ and
$\Delta\sigma_{xy}$ with the changing temperature and magnetic field
not match to the diffusion interaction correction. It is not surprising
because the parameter $T\tau=0.08-0.25$ is not small within this
temperature range and the ballistic part of interaction correction
gives significant contribution.

As we mention just below Eq.~(\ref{eq25}) the interaction contribution
can be extracted in this situation by analyzing the
$\Delta\sigma_{xx}$~vs~$T$ behavior at $B=1/\mu$. Such the dependence
plotted in Fig.~\ref{f1}(c) shows that the temperature dependence of
$\Delta\sigma_{xx}$ is really close to the logarithmic one and its
slope gives $K_{ee}=0.55 \pm 0.05$.

The diffusion part of the interaction correction has to lead to the
temperature dependence of the Hall coefficient, $\Delta R_H/R_H\simeq
-2\Delta\sigma_{xx}/\sigma_{xx}$, and, hence, the diffusion
contribution can be independently obtained from the $T$ dependence of
the Hall coefficient. As seen from Fig.~\ref{f1}(c) $\Delta\sigma_{xx}$
found as $[R_H(T)-R_H(4.2\text{ K})]\sigma_{xx}(4.2\text{
K})/2R_H(4.2\text{ K})$ agrees well with the data obtained by the first
method.

\begin{figure}
\includegraphics[width=\linewidth,clip=true]{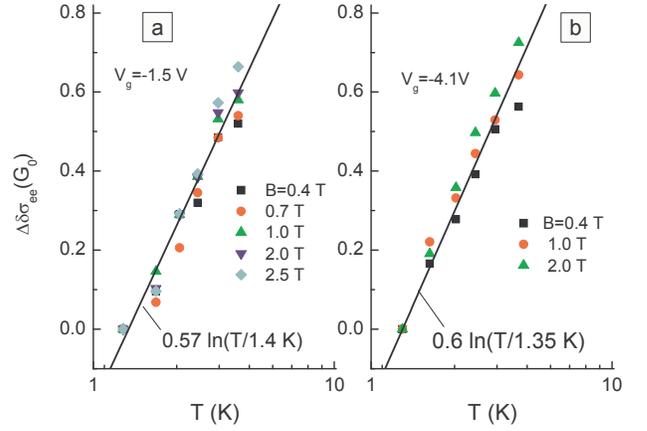}
\caption{(Color online) The temperature dependence of the diffusion
{\it e-e} interaction correction $\Delta\delta\sigma_{ee}=\delta\sigma_{ee}(T)-
\delta\sigma_{ee}(1.35\text{ K})$ for different magnetic fields near the balance
(a) and under the condition when
only lower quantum well is occupied (b). }\label{f2}
\end{figure}

Finally, the diffusion contribution $\delta\sigma_{ee}$ can be obtained
even over the whole magnetic field range by eliminating the ballistic
part of interaction with the use of the method described in
Ref.~\onlinecite{Min03-2}. Because the ballistic part of the
interaction correction is reduced to the renormalization of the
mobility and the diffusion part of the correction does not contribute
to the off-diagonal component of the conductivity, one can obtain the
$\mu$ vs $T$ dependence from $\sigma_{xy}$ knowing the electron density
(from the period Shubnikov-de Haas oscillations)
\begin{equation}
 \mu(T)=\left\{\frac{\sigma_{xy}(T)}{\left[en-\sigma_{xy}(T)B\right]B}\right\}^{1/2}
 \label{eq30}
\end{equation}
and find the correction $\delta\sigma_{ee}(T)$ as the difference
between the experimental value of $\sigma_{xx}(T)$ and the value of
$en\mu(T)/(1+\mu^2(T)B^2)$. The results of such a data treatment are
presented in Fig.~\ref{f2}(a) as the
$\Delta\delta\sigma_{ee}$~vs~$\ln{T}$ plot, where
$\Delta\delta\sigma_{ee}=\delta\sigma_{ee}(T)-
\delta\sigma_{ee}(1.35\text{ K})$. It is clearly seen that the slopes
of $\sigma_{xx}$~vs~$\ln{T}$ dependences are practically independent of
the magnetic field  and give $K_{ee}=0.57\pm 0.05$ that agrees well
with the value of $K_{ee}$ obtained by the two previous methods.

Thus, three different methods for obtaining of the diffusion
interaction correction give the same results. The correction
$\delta\sigma_{ee}$ is logarithmic in the temperature, and the value of
the parameter $K_{ee}$ is  $0.57\pm 0.05$.

Let us inspect the data for the case when only one well is occupied.
For the structure 3243 it occurs at $V_g\lesssim -4$~V. There are not
additional difficulties in the extraction of $K_{ee}$ for this case.
All three methods give also the same results. As an example we have
plotted in Fig.~\ref{f2}(b) the temperature dependence of
$\Delta\delta\sigma_{ee}$ taken at different magnetic field at
$V_g=-4.1$~V when the electron density and mobility are close to those
for each well at the balance. One can see that the temperature
dependences of $\Delta\delta\sigma_{ee}$ taken at different $B$ for
this case are close to each other also. The slope of the
$\Delta\delta\sigma_{ee}$~vs~$\ln{T}$ dependence gives $K_{ee}=0.60\pm
0.05$ that corresponds to $F_0^\sigma=-0.225$.  This value  is in a
good agreement with the theoretical estimate
$F_0^\sigma=-0.237$.\cite{Zala01}

The surprising thing is that the value of $K_{ee}$ in the balance
practically coincides with that for the regime when only one quantum
well is occupied. Such coincidence seems strange. It does not agree
with both cases discussed qualitatively in Section~\ref{sec:intr}.

\begin{figure}
\includegraphics[width=0.8\linewidth,clip=true]{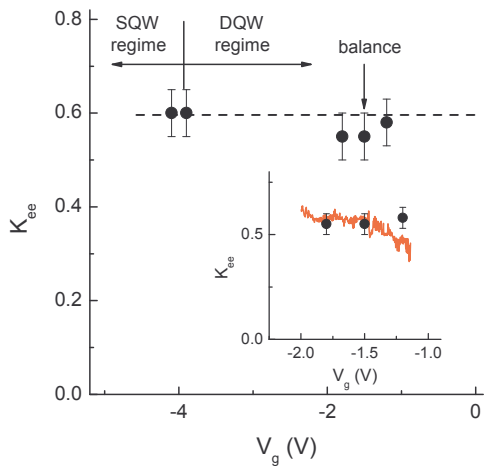}
\caption{(Color online) The  $K_{ee}$ values plotted against the gate voltage
for two regimes, when lower quantum well is occupied and near the balance.
In the inset, the line is the $K_{ee}$~vs~$V_g$ dependence obtained
when sweeping the gate voltage at different temperatures. }\label{f3}
\end{figure}

It is possible that the structure 3243 at $V_g=-1.5$~V is close to the
balance but not exactly in it. Let us analyze the data at the gate
voltages in the vicinity of $-1.5$~V. In this situation the mobilities
in the wells are distinguished and strictly speaking the method used
for $V_g=-1.5$~V is not applicable. However, one can easily assure that
by using  the total electron density $n_1+n_2$ and the average mobility
$\mu^*=\sigma^*/e(n_1+n_2)$ (where $\sigma^*=1/\rho_{xx}$ at
$B=1/\mu^*$) in the data processing one obtains the value of $K_{ee}$
very close to its average value. The results obtained by this way and
presented in Fig.~\ref{f3} show that the found values of $K_{ee}$ are
close to that for the balance.

The resonant change of $K_{ee}$ occurs quite possible within a very
narrow range of $V_g$ and we could overlook  it measuring $K_{ee}$ at
fixed $V_g$. To check such an occasion we have measured $\rho_{xx}$ and
$\rho_{xy}$ at fixed magnetic field $B=1/\mu_{b}$, where $\mu_{b}$
stands for the mobility in the balance, and different temperatures
sweeping the gate voltage. The dependence of $K_{ee}$~vs~$V_g$ was
found as the slope of the $\sigma_{xx}$ vs $\ln{T}$ plot. As seen from
the inset in Fig.~\ref{f3} $K_{ee}$ changes monotonically\footnote{The
monotonic run of experimental plot in inset in Fig.~\ref{f3} results
from the fact that the measurements were performed at fixed magnetic
field corresponding to $1/\mu_b$, whereas the mobilities at $V_g$ above
and below the balance gate voltage are somewhat different.} and
exhibits no resonant feature within the sweeping $V_g$ range. Analogous
results were obtained for the structure 3154.

Thus, all the results presented above show that the noticeable resonant
change of $K_{ee}$ to say nothing of change of its sign is not observed
in the structures investigated.

One possible reason of the absence of the $K_{ee}$ resonance can be the
fact that the scatterers are not common enough in spite of  our efforts
to design special structures and despite proximity of the mobilities in
the wells. The unavoidable variation in the scatterers positions with
respect to the center of the barrier results in the fact that the
specific impurity scatters the carriers of the lower and upper wells
differently. Besides, the interwell distance is not sufficiently small
in our case. The parameter $\varkappa d$  is equal to $3.6$, so the
interaction between the electrons in the different wells is noticeably
weaker than that between electrons within the one well.

On the other hand, the interaction correction for the structure with
the large interwell distance ($ d \gg 1/\varkappa$) should be equal to
the sum of the correction in wells. The value  of $K_{ee}$ for the case
when one well is occupied is  $0.60\pm 0.05$ for structure 3243 (see
Table~\ref{tab1}). So, at $V_g=-1.5$~V, when both wells are occupied
and each of them has approximately the same density, the value of
$K_{ee}$ would be expected as large as $\simeq 1.2$. In reality the
$K_{ee}$ value is twice less. We conceive that this paradox can be
resolved by taking into account the screening of the {\it e-e}
interaction between the carriers in the one well by the carriers of the
other one, which was not taken into account at qualitative
consideration above. This screening reduces the {\it e-e} interaction
strength and, consequently, diminishes the interaction contribution to
the conductivity.
\begin{figure}
\includegraphics[width=0.8\linewidth,clip=true]{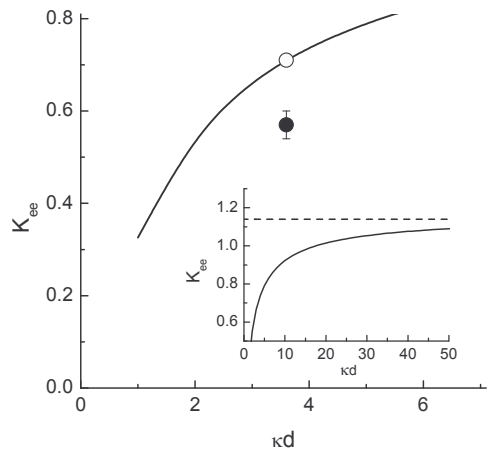}
\caption{(Color online) The $K_{ee}$ value at the balance plotted as a function of $\varkappa d$.
The solid line  is calculated according to Ref.~\onlinecite{Burmistrov2010}
with $n=7.5\times
10^{11}$~cm$^{-2}$ that corresponds to the structure 3243 in the balance.
The solid circle is obtained experimentally, the open circles marks
the theoretical value of $K_{ee}$ for the structure 3243.  The inset shows the calculated
$K_{ee}$~vs~$\varkappa d$ dependence in wider $\kappa d$ range (solid line).  The dashed
line is the $K_{ee}$ value in the limiting case of two independent wells. }\label{f4}
\end{figure}

It is clear that for the adequate understanding of the role of the {\it
e-e} interaction in the double well structures, the theory, which
properly takes into account the interaction in the singlet and
multiplet channels and specifics of the screening for different
interwell distances, is necessary. Such theoretical consideration is
presented in Ref.~\onlinecite{Burmistrov2010}. The authors analyze both
the interaction and weak localization corrections to the conductivity
of the double layer structures in framework of the random phase
approximation. The interaction effect is considered for the case of the
identical layers. Unexpected result is that the multiplet contribution
even in the case of common scatterers does not win the singlet
contribution for actual parameter $\varkappa d >1$ and, consequently,
does not result in the change of the $K_{ee}$ sign, as we have naively
reasoned in the Section~\ref{sec:intr}.

Let us compare the theoretical results\cite{Burmistrov2010} with the
experimental data. For both structures 3243 and 3154, the experimental
and theoretical values of $K_{ee}$ for SQW regime and for the balance
are presented in Table~\ref{tab1}. The calculations have been performed
with the electron densities listed in the table,
$\varkappa=2\times10^6$~cm$^{-1}$, and $d=1.8\times 10^{-6}$~cm.
Because no fitting parameters have been used, agreement between the
theory and experiment can be considered as reasonably good.

Figure~\ref{f4} illustrates the sensitivity of $K_{ee}$ to the
interwell distance. The theoretical and experimental values of $K_{ee}$
corresponding to the sample 3243 are marked by open and solid circles,
respectively.  Inspection of this figure shows that contrary to our
expectation the value of $K_{ee}$ does not change the sign at any real
interwell distances, all the more it does not acquire the value
$K_{ee}\simeq-1$, which corresponds to the equal contributions of $15$
multiplet channels. Besides, the other limiting case of independent
contributions coming from each well is achieved at very large distance,
$\varkappa d>30$ (see inset in Fig.~\ref{f4}). This is clear indication
of great importance of the specific feature of the screening of
electron-electron interaction in the double layer systems.

\section{Conclusion}

We have studied the electron-electron interaction  correction to the
conductivity of 2D electron gas in the gated double quantum well
Al$_x$Ga$_{1-x}$As/GaAs heterostructures. Using three different methods
we have obtained the diffusion part of the interaction correction under
the conditions when one and two quantum wells are occupied. It has been
found that the interaction correction, contrary to naive expectations,
is practically independent of whether two or one quantum well
contribute to the conductivity. This observation is consistent with the
results of the paper by Burmistrov, Gornyi and
Tikhonov,\cite{Burmistrov2010} in which the theory for the dephasing
and electron-electron interaction in the double well structures is
developed.

\section*{Acknowledgments} We would like to thank I.~S.~Burmistrov and
I.~V.~Gornyi for illuminating discussions. This work has been supported
in part by the RFBR (Grant Nos 09-02-12206, 09-02-00789, 10-02-91336,
and 10-02-00481).

% \bibliography{QuantumCorrections}

%merlin.mbs 2010-03-15 4.21a (PWD, AO, DPC)

\end{document}